# Genetic control and geo-climate adaptation of pod dehiscence provide novel insights into the soybean domestication and expansion

Jiaoping Zhang* and Asheesh K. Singh*
Department of Agronomy, Iowa State University, Ames, IA 50011, USA
* Correspondence: singhak@iastate.edu, jiaoping@iastate.edu

## ABSTRACT

Loss of pod dehiscence is a key step during soybean [*Glycine max* (L.) Merr.] domestication. Genome-wide association analysis for soybean shattering identified loci harboring *Pdh1*, *NST1A* and *SHAT1-5*. Pairwise epistatic interactions were observed, and the dehiscent *Pdh1* overcomes the resistance conferred by *NST1A* or *SHAT1-5* locus, indicating that *Pdh1* predominates pod dehiscence expression. Further candidate gene association analysis identified a nonsense mutation in *NST1A* associated with pod dehiscence. Allele composition and population differential analyses unraveled that *Pdh1* and *NST1A*, but not *SHAT1-5*, underwent domestication and modern breeding selections. Geographic analysis showed that in Northeast China (NEC), indehiscence at both *Pdh1 and NST1A* were required by cultivated soybean; while indehiscent *Pdh1* alone is capable of coping shattering in Huang-Huai-Hai (HHH) valleys where it originated; and no specific indehiscence was required in Southern China (SC). Geo-climatic investigation revealed strong correlation between relative humidity and frequency of indehiscent *Pdh1* across China. This study demonstrates that the epistatic interaction between *Pdh1* and *NST1A* fulfills a pivotal role in determining the level of resistance against pod dehiscence. Humidity shapes the distribution of indehiscent alleles. Our results also suggest that HHH valleys, not NEC, was at least one of the origin centers of cultivated soybean.

**Key words:** soybean, domestication, pod dehiscence, seed shattering, candidate gene association analysis, geo-climate adaptation.

## INTRODUCTION

Modern crop species have undergone domestication that makes them distinct from their wild ancestors. During domestication, the fitness of a plant for human exploitation increases through artificial selection for a suite of traits including seed shattering (or pod dehiscence), seed size, branching and stature (Meyer *et al.* 2012, Meyer Purugganan 2013). Seed shattering is pivotal for the propagation of wild plant species. However, it is unfavorable for crop production, because it causes yield loss prior to harvesting. The elimination of seed shattering is vital for seed retention and is a key step during crop domestication.

The genetic mechanism underlying seed shattering varies across species. In monocot crops, such as cereals, the abscission layer between the hull and pedicle is necessary for seed shattering. Genes *SH4* (allelic to *SHA1*) (Li *et al.* 2006, Lin *et al.* 2007), *qSH1* (Konishi *et al.* 2006), *Sh1* (Lin *et al.* 2012), *SHAT1 and SH5* (Zhou *et al.* 2012, Yoon *et al.* 2014) that regulate the development of the abscission layer are responsible for seed shattering in rice (*Oryza sativa* L.). *Sh1* was also revealed under parallel domestication in sorghum (*Sorghum bicolor* (L.) Moench), rice (*O. sativa*), and maize (*Zea mays* L.) (Lin *et al.* 2012). However, in dicot crops, such as soybean [*Glycine max* (L.) Merr.], the abscission layer remain unchanged between the wild ancestors (*G. soja*) and domesticated soybeans (Dong *et al.* 2014), indicating new strategies of pod indehiscence in soybean that is different from cereals. Recent studies identified two genes controlling soybean pod dehiscence via distinct mechanisms. The *Pdh1* encodes a dirigent family protein, which is highly expressed in the lignin-rich inner sclerenchyma of pod walls. The functional *Pdh1* coils pod walls of mature plants under low humidity conditions and serves as a driving force for pod dehiscence (Funatsuki *et al.* 2014). *SHAT1-5*, an *NAC* gene, conditions the deposition of the secondary walls of the lignified fiber cap cells (FCC) in the pod ventral suture and determines the binding strength of the pods (Dong *et al.* 2014). Furthermore, genetic mapping studies identified additional quantitative trait loci (QTL) associated with pod dehiscence (SoyBase, https://soybase.org/), implied a complex genetic regulatory network of pod dehiscence in soybean.

Understanding the geo-climatic adaptation of crop species is a pressing need in order to develop resilient cultivars and ensure food security under changing climates. A recent study illustrated that environment accounts for a substantial portion of genetic variation (Lasky *et al.* 2015). Both wild and cultivated soybeans exhibited ecological differentiation related to geographic conditions (Ding *et al.* 2008). Based on the topographic distribution and the soybean growth habit, the soybean-growing areas in China can be divided into three zones: the Northeast China (NEC), the Huang-Huai-Hai (HHH) valleys and the Southern China (SC) (Ding *et al.* 2008). The investigation of the geographic adaptation of the pod dehiscence will provide insights into the domestication and improvement of soybean. Previous study indicated that humidity is a crucial factor in pod dehiscence in soybean (Tsuchiya 1987). However, the impact of climate conditions on the genetic architecture of this important soybean domestication trait remains unclear.

Here, we report a novel locus, *NST1A*, and the known *Pdh1* and *SHAT1-5* loci associated with soybean pod dehiscence through genome-wide association study (GWAS). The causal genetic variants were further verified through candidate gene association analysis. We revealed the epistatic interaction between these loci, and demonstrated its importance to soybean domestication, especially *Pdh1* and *NST1A*. This study uncovered the role of humidity on shaping the distribution of the pod indehiscent alleles across regions in China as part of genome-environment adaptation during soybean domestication. It also provides insights into the origin centers and expansion of cultivated soybean.

## MATERIALS AND METHODS

### Plant materials and phenotyping

The pod shattering dataset of the study SOYBEAN.EVALUATION.MS923 was retrieved from Germplasm Resources Information Network (GRIN, http://www.ars-grin.gov/), which consists of 782 soybean plant introductions (*G.max*) of the USDA soybean germplasm collection belonging to maturity group VI. These accessions were planted at Stoneville, Mississippi, in 1992 and 1993 with one replication each year, and the average values were used for analysis. The details of experimental design and trait phenotyping are described in a previous study (Hill *et al.* 2001). Briefly, plots were 4-rows wide, with rows 3.6 m long and 91 cm row spacing. The border rows were evaluated for pod dehiscence two weeks after the center two rows were harvested using a scale of 1 - 5 based on the percentage of open pods: 1 = 0%; 2 = 1 - 10%; 3 = 10 - 25%; 4 = 25 - 50% and 5 = over 50% shattering.

## Genotyping and association analysis

The single nucleotide polymorphism (SNP) dataset for the association panel, prepared by using the SoySNP50K Illumina Infinium BeadChip (Song *et al.* 2013), was retrieved from SoyBase (https://soybase.org/). The quality control and imputation of missing data were as described in a previous study (Zhang *et al.* 2015). Finally, 30,530 SNPs with minor allele frequencies > 5% remained for further study. The association analysis with mixed linear model and general liner model were conducted by using the Genome Association and Prediction Integrated Tool (GAPIT) software implemented in R as previously described (Zhang *et al.* 2010, Lipka *et al.* 2012). No correction of population structure was suggested according to the Bayesian Information Criterion test output by GAPIT. Additionally association analysis with the first three principal components accounting for population structure also gave similar results. The significant threshold was corrected for multiple testing by using the Bonferroni correction ($\alpha = 0.05$).

## Allele distribution and humidity

A total of 758 *G.soja* from China, Koreas and Japan, 13,371 *G.max* Asian landraces and 834 North American cultivars (after removing the isolines) deposited in GRIN were involved in the allele distribution analysis. The SNP data of these soybean panels was retrieved from SoyBase (https://soybase.org/). The origin information of the accessions was available on GRIN. The map was created using the R 'maps' package (Team 2012, Becker *et al.* 2013). Data for the relative humidity at 10 m above the surface of the earth and air temperature were obtained from NASA surface meteorology and solar energy (release 6.0) (https://eosweb.larc.nasa.gov/cgi-bin/sse/global.cgi?email=skip@larc.nasa.gov). The dataset contains monthly and annual average relative humidity and air temperature from July 1983 to June 2005 on one-by-one latitude/longitude degree resolution. Only the average humidity and temperature values across the soybean harvest season including September, October, and November during these 22 years were used for analysis.

## Population differential analysis

A total 153 *G.soja* originated from China and the same number of Chinese landraces (*G.max*) randomly selected from NEC, HHH valleys or SC were used to calculate the population differential ($F_{st}$) of the specific chromosomal regions with the R 'snpStats' package (Clayton 2012). A diverse *G.max* panel that consists of equal number of accessions (n=51) randomly selected from NEC, HHH valleys and SC was also used for population differential analysis.

## Alignment and phylogenetic analysis of Pdh1, NST1A and SHAT1-5 proteins

Because *NST1A* and *SHAT1-5* are paralogs and highly similar (Dong *et al.* 2013), only amino acid sequence of *NSAT1A* and *Pdh1* of soybean was searched from homologs using BLASTP against the entire GenBank on the National Center of Biotechnology Information (NCBI, https://www.ncbi.nlm.nih.gov/). The top hits were selected and the alignment analysis was conducted on NCBI. The phylogenetic analysis of the aligned proteins was carried out using the R 'ape' package (Paradis *et al.* 2004).

## Candidate gene association study

A subset of 400 accessions was randomly selected from the original association panel for candidate gene association analysis. Twenty-one SNPs located at the promoter and the coding regions of the candidate gene *Glyma07g05660*, and another 10 SNPs and one InDel including the known causal genetic variants of *Pdh1* and *SHAT1-5* were selected based on the SNP dataset on Phytozome (https://phytozome.jgi.doe.gov) (Zhou *et al.* 2015). The genomic DNA preparation and SNP genotyping were conducted by the LGC Genomics at Beverly, MA. The primers for the InDel and the polymerase chain reaction (PCR) conditions were listed in Table S1. The sequencing validation of the InDel polymorphism was conducted as described previously (Xu *et al.* 2015). The PCR products and sequencing results are shown in Fig. S1. The association analysis was implemented in GAPIT as described above with a significance threshold $P < 10^{-3}$.

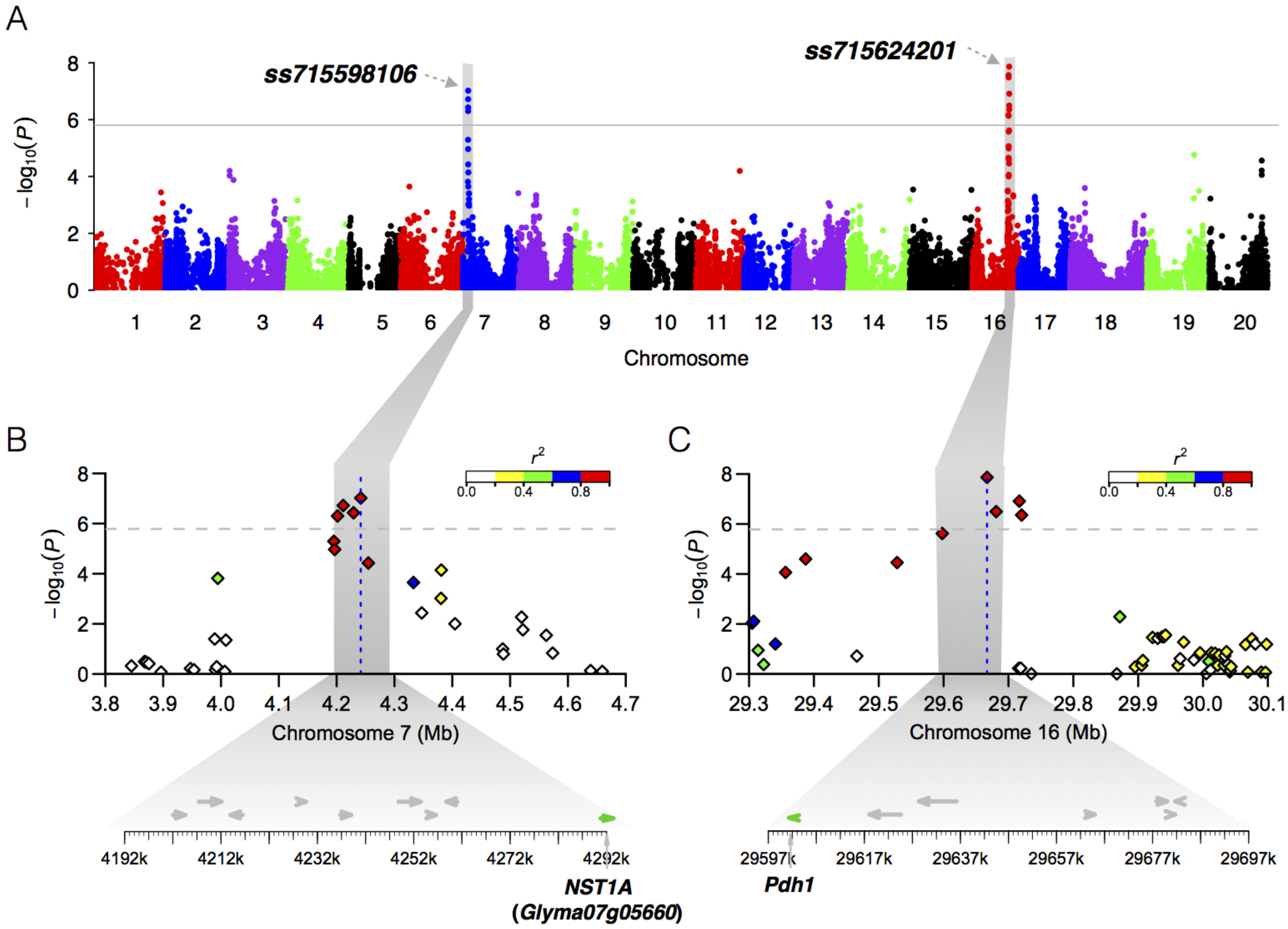

**Fig 1. Manhattan plots and candidate genes of the loci associated with pod dehiscence in soybean.** (A) Negative $\log_{10}$-transformed *P* values from a genome-wide scan using mixed linear model are plotted against positions on each of 20 chromosomes. The grey line indicates the significance threshold ($P = 1.64 \times 10^{-6}$). The lead SNP of each locus is given. (B) and (C) Candidate genes of the loci associated with pod dehiscence. The top panel shows regional Manhattan plot for the indicated region. The color of each SNP indicates its linkage disequilibrium $r^2$ value with the peak SNP as shown in the color intensity index on the top-right. The bottom panel shows all putative genes in the indicated region as indicated by the shadow.

## RESULTS

### GWAS identified loci associated with pod dehiscence

Genome-wide scan with the mixed linear model (MLM) identified two QTL, the SNP *ss715598106* on Gm07 (also known as *Gm07_4241705_G_T*) and *ss715624201* on Gm16 (also known as *Gm16_29666971_T_C*), strongly associated with pod dehiscence (Fig. 1A). The locus tagged by *ss715598106* on Gm07 has not been reported previously. A closer review of the region identified the candidate gene *NST1A* (*Glyma07g05660*), that is 49 kb downstream of *ss715598106* and encodes a No Apical Meristem (NAM) protein (Fig. 1B). *NST1A* is a paralogue of *SHAT1-5* (also known as *NST1B*). They shares 92.8% amino acid similarity and similar expression profiles (Dong *et al.* 2013). On Gm16, *ss715624201* is in a hotspot associated with soybean pod dehiscence (Bailey *et al.* 1997, Funatsuki *et al.* 2006, Kang *et al.* 2009), and is 65 kb upstream of the shattering gene *Pdh1* that was characterized in a recent study (Fig. 1C) (Funatsuki *et al.* 2014). *SHAT1-5* locus was not detected using MLM, even with specifying *ss715598106* and/or *ss715624201* as covariate. However, using general linear model (GLM) without correction of kinship, *ss715623567* ($P$ = 3.8 x $10^{-11}$) located at 581 bp upstream of *SHAT1-5* and *ss715624201* were significant, but *ss715598106* was not significant.

### Epistatic interaction determines the level of resistance to pod dehiscence

Pairwise interactions among three loci were detected, but three-way interactions were non-significant (Table S2). The strong epistatic interaction between *ss715624201* and *ss715598106* ($P$ = 1.3 x $10^{-5}$) was noted, and at least partially explains why *ss715598106* was not detected in the GLM. Under the condition of dehiscence genotype (‘*CC*’) at *ss715624201*, the change of genotype at

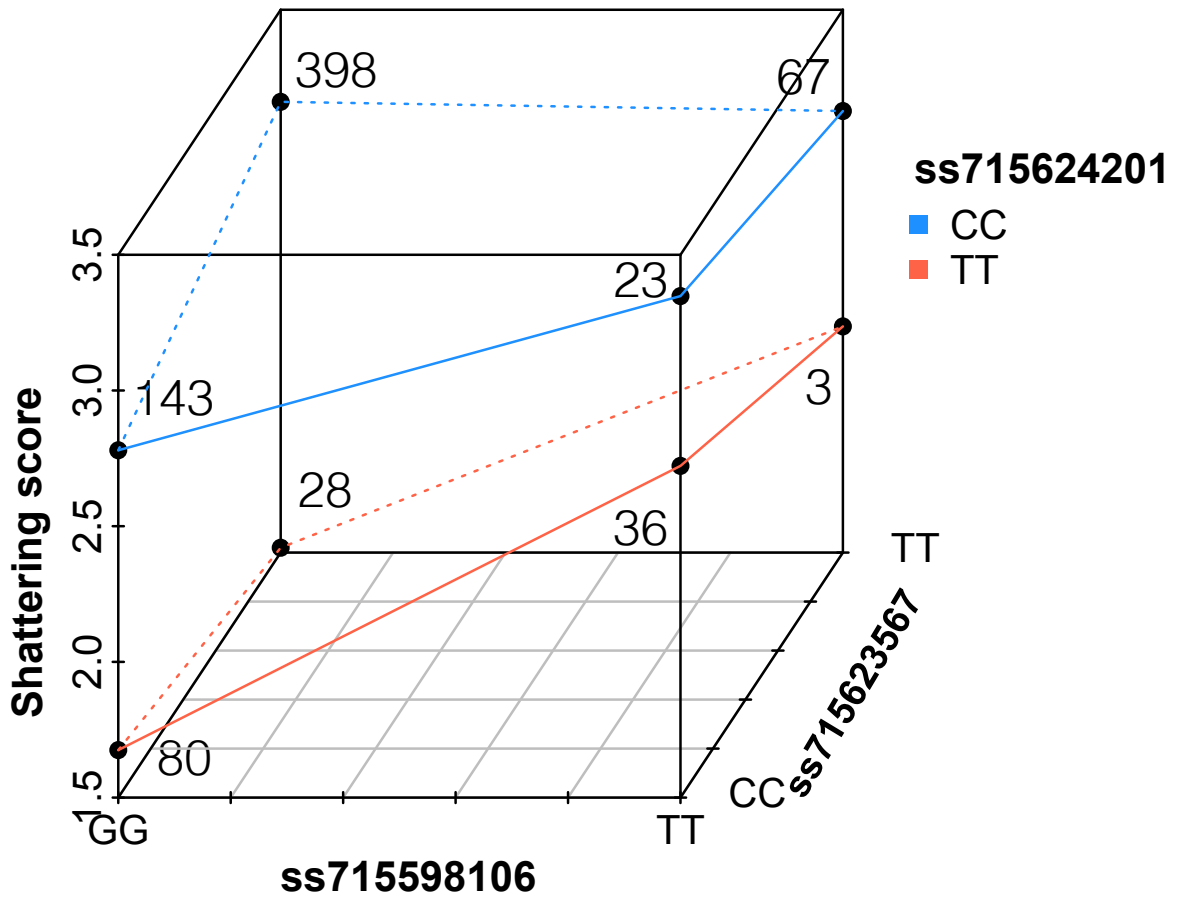

**Fig 2 Epistatic effect between the loci associated with pod dehiscence in soybean.** Each dot indicates the average shattering score of the relevant genotypic group. The number of accessions of each genotype group is also given.

*ss715598106* or *ss715623567* has very limited impact on trait performance (Fig. 2) suggesting that *ss715624201* locus (harboring *Pdh1*) dominates pod dehiscence and is able to overcome the indehiscence conferred by *ss715598106* (harboring *NST1A*) and *ss715623567* (carrying *SHAT1-5*). This is consistent with the role of the related genes in pod dehiscence. In soybean, *Pdh1* is associated with the coiling of the drying pod wall and servers as the driving force for dehiscence under low humidity condition (Funatsuki *et al.* 2014). *NST1A* and *SHAT1-5* are paralogs (Dong *et al.* 2013), and *SHAT1-5* has been shown to thicken the FCC secondary walls that is associated with binding strength between pod walls (Dong *et al.* 2014). At the indehiscent (‘*TT*’) background of *ss715624201*, switching genotypes of *ss715598106* from ‘*TT*’ *to* ‘*GG*’ or *ss715623567* from *‘CC’* to ‘*TT*’, strengthens the dehiscent reistance indicating that ‘*G*’ and ‘*T*’ are the indehiscent alleles, respectively. The results also illustrated that the resistance conferred by *ss715624201* and *ss715598106* is comparable to that from all three loci combined (Fig. 2). However, it is difficult to distinguish the effect of *ss715598106* and *ss715623567* due to lack of enough (only three) individual with '*TT*' at *ss715623567* under the resistant background at *ss715624201* (Fig. 2).

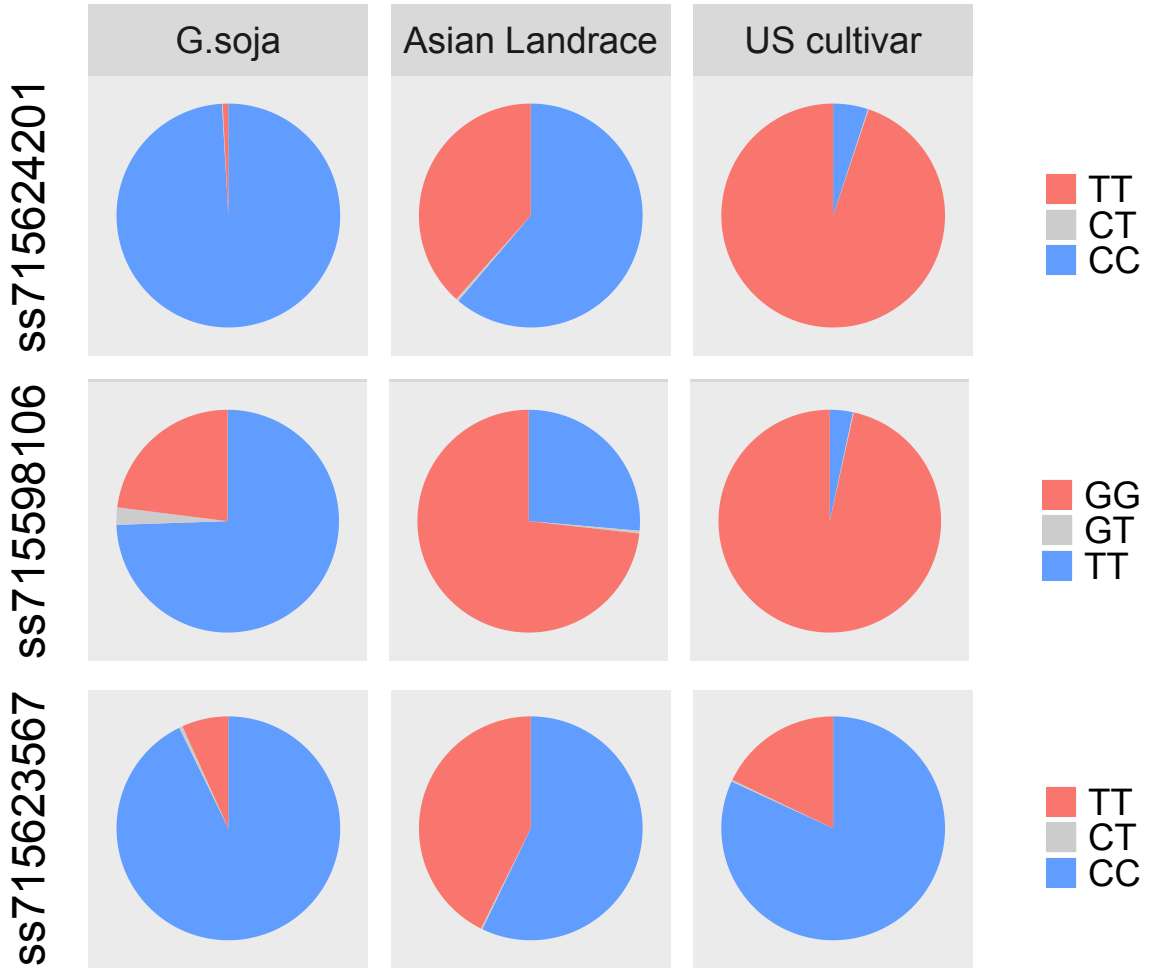

**Figure 3. Allele frequency of *ss715624201*, *ss715598106* and *ss715623567* associated with soybean pod dehiscence among *Glycine soja*, landraces, and modern cultivars.** A total of 758 *G.soja* originated from China, Koreas and Japan*;* 13,371 Asian soybean landraces (*G.max*); and 834 modern soybean cultivars released in North America are included in the analysis. The genotypic data of the two loci were retrieved from SoyBase (https://soybase.org/).

## The origin and transition of pod-indehiscent alleles during soybean domestication

Further investigation showed that the indehiscent alleles are minor at all three loci in wild soybean (Fig. 3), particularly *ss715624201* and *ss715623567*, indicating natural selection against pod indehiscence. The rare (frequency < 5%) indehiscent allele of *ss715624201* implied that it was under high natural selection pressure because of the large effect. The indehiscent allele frequencies of all three loci were substantially increased (landraces versus *G.soja*) during domestication. Although loss of pod dehiscence is important for soybean domestication, the indehiscent allele of the major-effect *ss715624201* remained minor within Asian landraces indicating that necessity of pod indehiscence for soybean domestication varies across different origin or environmental conditions. In North America cultivars, the indehiscent alleles at *ss715624201* and *ss715598106* continuously increased until being almost fixed, but that of *ss715623567* decreased and remained minor, suggesting *ss715623567* was not under modern breeding selection in the United States (US). Besides domestication, North American cultivars also underwent introduction bottleneck (Hyten *et al.* 2006). A survey of the 17 Asian landraces that account for 86% of the parentage of modern US cultivars indicated that the decrease of indehiscent *ss715623567* in cultivars might be attributed to the founder population effect (Table S3).

Geographic origin analysis of the *G.soja* accessions revealed that only six of the 758 wild accessions that originated from China, Koreas and Japan carried the ancient indehiscent allele of *ss715624201.* All of them were from China and were collected from HHH valleys (Fig. 4A and Table S4), which is considered one of the centers of origin of cultivated soybean (Zhou *et al.* 1998). An

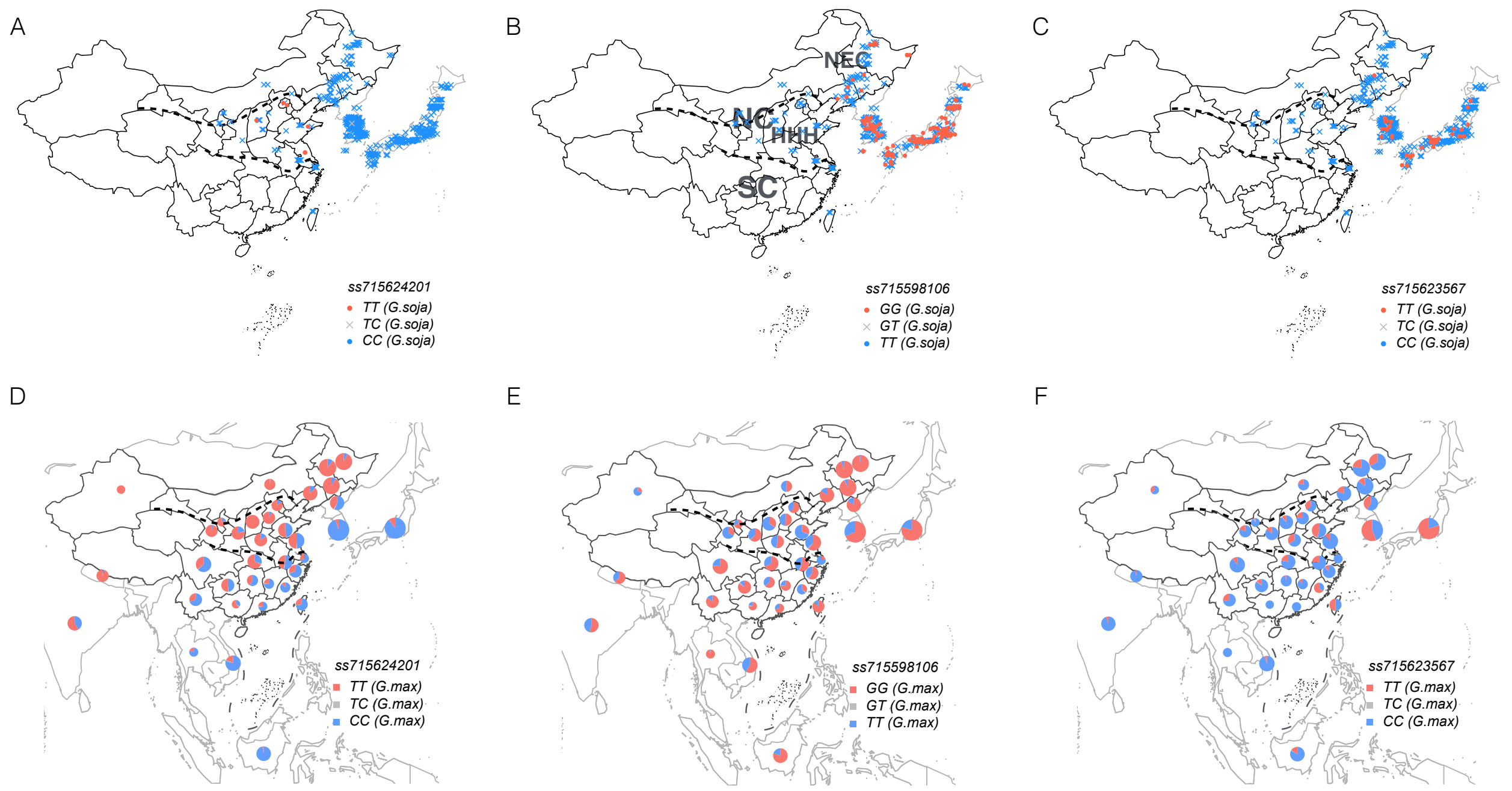

**Figure 4. Geographic distribution of the pod dehiscence alleles at *ss715624201*, *ss715598106* and *ss715623567* in wild soybean and Asian landraces.** Shown are allele distributions of relevant loci within 758 wild soybean accessions (*G.soja*) across China, Koreas and Japan (A-C) and the Asian landraces (*G.max*) (D-F) based on their origin. A total of 12,441 soybean landraces that with known origin and each origin has ≥ 20 accessions are plotted. Among them, 721 accessions from Northeast China without knowing the specific origin province are also plotted. The radius of the pie indicates one half of the $\log_{10}$-transformed number of germplasm accessions of each origin. Bottom dashed line is used to delineate Northern China (NC) with Southern China (SC), while Huang-Huai-Hai (HHH) valleys is the region between bottom and top dashed lines.

investigation of a previous resequencing study identified 12 wild soybeans originated from SC (Zhejiang Province) that were not in the USDA germplasm collection (Zhou *et al.* 2015). None of them carried the indehiscent allele of *ss715624201*. Interestingly, the allele distribution of *ss715624201* in landraces showed a clear pattern where the indehiscent allele frequency increased from SC to Northern China (NC), and from the coastland (including Japan, Korea and Indonesia) to inland regions (Fig. 4D). In the wild progenitors, *ss715598106* and *ss715623567* showed similar distribution. The indehiscent alleles of both loci were mainly located in Koreas and Japan, and few in NEC (Fig. 4B and C). In the landraces, notably, the indehiscent allele of *ss715598106* was predominant among the accessions from SC and NEC, but not HHH valleys (Fig. 4E). However, no special pattern was found for *ss715623567* (Fig. 4F). Above results suggest that the pod indehiscence conferred by *ss715624201* alone may not be strong enough to prevent shattering in NEC; and *ss715598106*, instead of *ss715623567,* was selected to enhance the resistance to pod dehiscence during soybean domestication. At HHH valleys, indehiscent *ss715624201* alone was capable of coping with shattering. At SC, resistance to dehiscence might not be required for cultivated soybean given that indehiscent

*ss715624201* was minor in SC but it was required for resistance of other two loci as illustrated by the epistatic effects.

We further estimated the $F_{st}$ between soybean landraces and the wild relatives from different regions in China at the candidate chromosomal region of the three loci. Consistent with the allele distribution, the landraces in NEC gave the largest differential from the wild progenitors at regions of both *ss715624201* and *ss715598106* loci, followed by the HHH panel of landraces at *ss715624201* and combined panel at *ss715598106* (Fig. 5A-B). No selection trace was observed at *ss715623567* across different panels (Fig. 5C). Combined panel of landraces from multiple regions in China was widely used in a genome-wide scan for domestication related sweeps in soybean (Zhou *et al.* 2015, Han *et al.* 2016, Wang *et al.* 2016). Although a similar region on Gm16 was detected, *Pdh1* locus was not detected in a recent study (Zhou *et al.* 2015). These results suggested that using domestication center specific population of landraces (for example, NEC, SC, HHH), instead of combined population with pooled multiple origins, might improve the power to detect the domestication related chromosomal regions and generate insights on domestication.

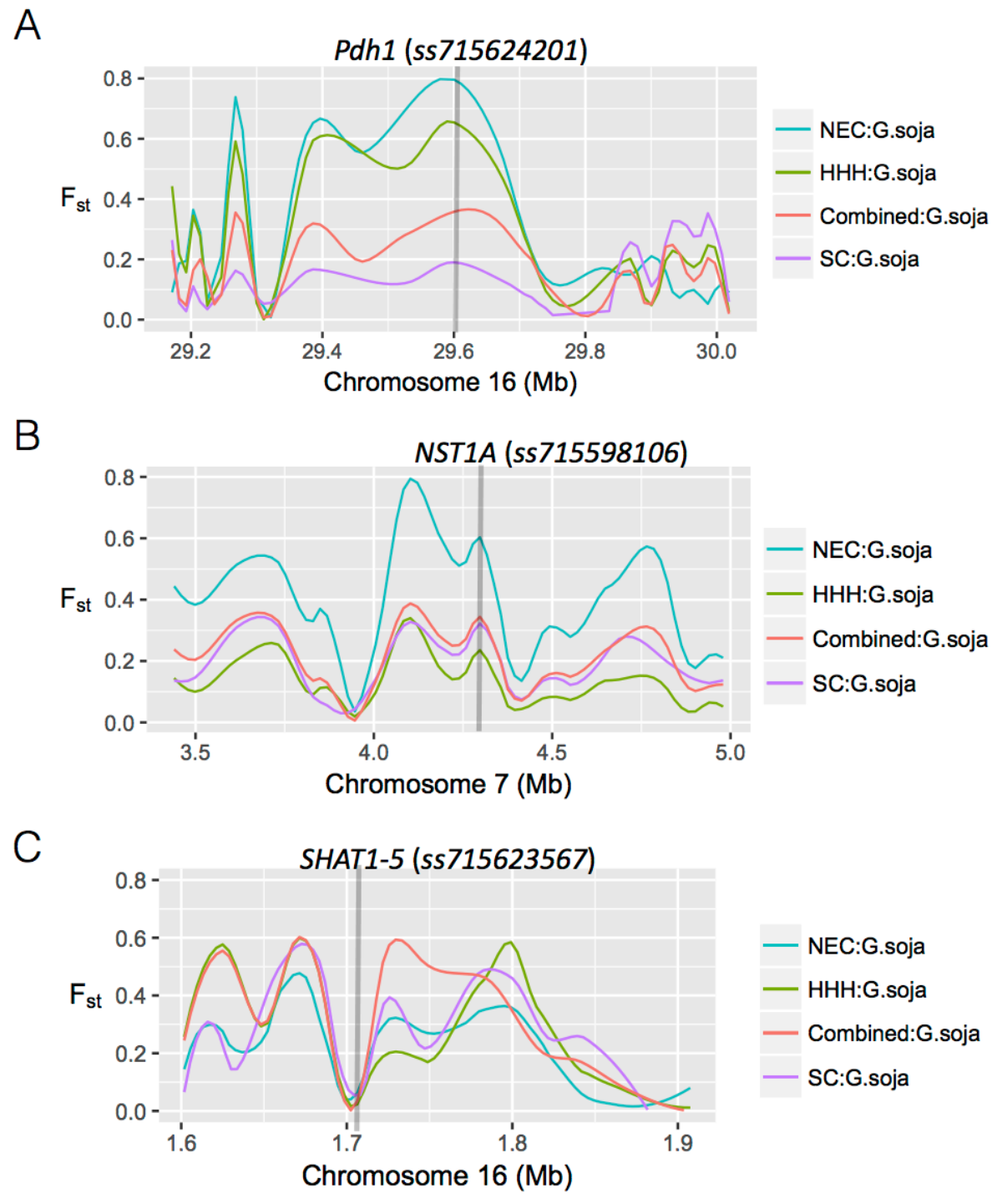

**Figure 5. Population differential ($F_{st}$) at the chromosomal regions associated with pod dehiscence.** (A), (B) and (C) $F_{st}$ values of the SNPs within the indicated chromosomal regions tagged by *ss715624201*, *ss715598106* and *ss715623567* respectively, between soybean landraces (*Glycine max*) and wild relatives (*G.soja*). A total of 153 Chinese wild soybean and 153 randomly selected soybean landraces originated from Northeast China (NEC), the Huang-Huai-Hai valleys (HHH), Southern China (SC) or combined panel are used for analysis. The combined panel consists equal number of accessions randomly selected from NEC, HHH valleys and SC. The vertical lines indicate the location of the candidate genes.

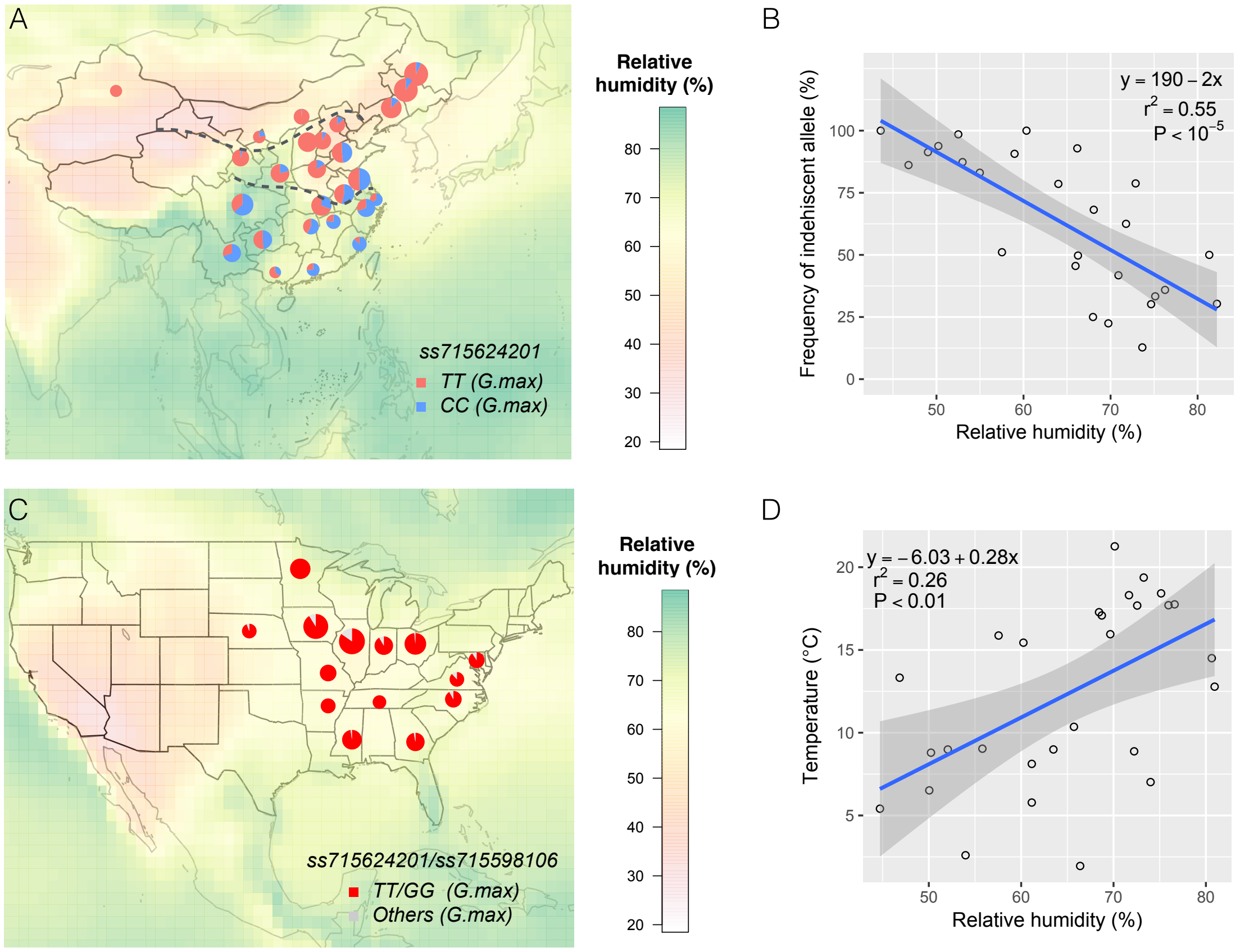

**Figure 6. Geo-climate distribution of the pod dehiscent alleles, and the correlations between the relative humidity and the allele frequency of indehiscent *ss715624201*.** (A) Geo-climate distribution of *ss715624201* in Chinese landraces. Shown are 4,228 landraces from 26 provinces in China that each has ≥ 20 accessions. The radius of pie indicates the regional population size as described above. The heatmap indicates the average relative humidity across the soybean harvest season (September, October, and November) of the related regions in a resolution of one latitude/longitude degree from July 1983 to June 2005 (NASA, https://eosweb.larc.nasa.gov/). (B) The correlation between the relative humidity and the distribution of the indehiscent *ss715624201* among the landrace showing in (A). (C) Geo-climate distribution of *ss715624201* and *ss715598106* in US cultivars. A total 645 from 14 States with each having ≥ 10 cultivars are plotted. (D) The correlation between relative humidity and air temperature across China. The calculation is based on the average temperature and humidity values during the same time period as in (A).

## Relative humidity shapes the geographic distribution of the pod-dehiscence alleles

The strong geographic pattern of the allele distribution at *ss715624201* encouraged us to further explore the underlying driven force. Previous studies suggested that humidity is the major environmental factor affecting pod dehiscence in soybean (Tsuchiya 1987, Funatsuki *et al.* 2014). Therefore, we mapped the relative humidity on top of the allele distribution of the landraces across China (Fig. 6A). The results showed that the relative humidity increases from NEC, HHH valleys to SC region. This matches the change of the level of resistance to pod dehiscence very well. The NEC has low humidity and requires the highest level of pod-indehiscence at both *ss715624201* but *ss715598106*; the HHH valleys has moderate humidity and only requires the pod indehiscence conferred by major-effect *ss715624201*; while the broad SC area has high humidity and requires no indehiscence at *ss715624201* (Fig. 4 and 6). High correlation was discovered between the level of the humidity and the allele frequency for *ss715624201* ($r$ = -0.75, $P < 10^{-5}$, Fig. 6B) but not for *ss715598106* or their combination (data not shown), which supports the function of *Pdh1*, candidate gene for *ss715624201*, that conditions coiling force of drying pod walls. Surface humidity and temperature are highly correlated. According to a reconstruction study of regional and global temperature for the Holocene (Marcott *et al.* 2013), the past decades global temperatures were comparable to that of 6,000-9,000 years ago when soybean was believed to be domesticated (Carter *et al.* 2004). Therefore, above results demonstrate the importance of RH in shaping the geographic pattern of *ss715624201* allele distribution and the different levels of resistance to pod dehiscence across China during soybean domestication.

A survey of the soybean cultivars developed in the US showed that the indehiscent allele was predominant at both *ss715624201* and *ss715598106*, which obviously was not driven by humidity conditions alone (Fig. 6C). Other driving factors include modern breeding selection against shattering to reduce yield losses and the low genetic diversity of the US soybean that experienced multiple genetic bottlenecks (Gizlice *et al.* 1994, Hyten *et al.* 2006). If the latter is the case, the local moderate humidity level gives US soybean breeding programs great potential to enhance the genetic diversity by utilizing germplasm resources with less pod indehiscence. This potential increases under the context of global warming, because the dehiscent allele averagely decreases 14.3% for every 2°C increase during maturing season (Fig. 6B and D).

## Candidate gene association analysis identified premature stop mutations associated with pod indehiscence

Promising candidate genes inspire further investigation of the casual genetic variants. Non-synonymous SNPs associated with pod dehiscence were identified in *Pdh1* and *NST1A* through candidate gene association analyses using MLM (Fig. 7A-B). Both resulted in truncated transcriptions. The *ss.101224845* introduced a stop codon close to the N-terminus of Pdh1 accounting for pod indehiscence as reported in a recent study (Funatsuki *et al.* 2014). Additionally, *ss.101224850,* located at the promoter of *Pdh1* and in complete linkage disequilibrium (LD, $r^2$ = 1) with *ss.101224845,* was also identified. The *ss.98955957* introduced a premature stop codon close to the C-terminus of *NST1A* and led to a missing of 47 amino acids that related to dehiscence resistance. As expected, *ss.101224845* and *ss.98955957* were in high LD with *ss715624201* ($r^2$ = 0.83) and *ss715598106* ($r^2$ = 0.85) respectively. However, the previously reported causal genetic variant of pod dehiscence at the promoter of *SAHT1-5* (designated as *pShat1-5*) was only detected using GLM ($P$ = 7.2 x $10^{-6}$) (Fig. 7C). Similar to other two loci, *pShat1-5* was in high LD with *ss715623567* ($r^2$ = 0.82), which is only 3.4 kb apart. The known-indehiscent allele of *pShat1-5* and the allele '*T*' at *ss715623567* formed the major haplotype, suggesting that '*T*' is associated with pod indehiscence.

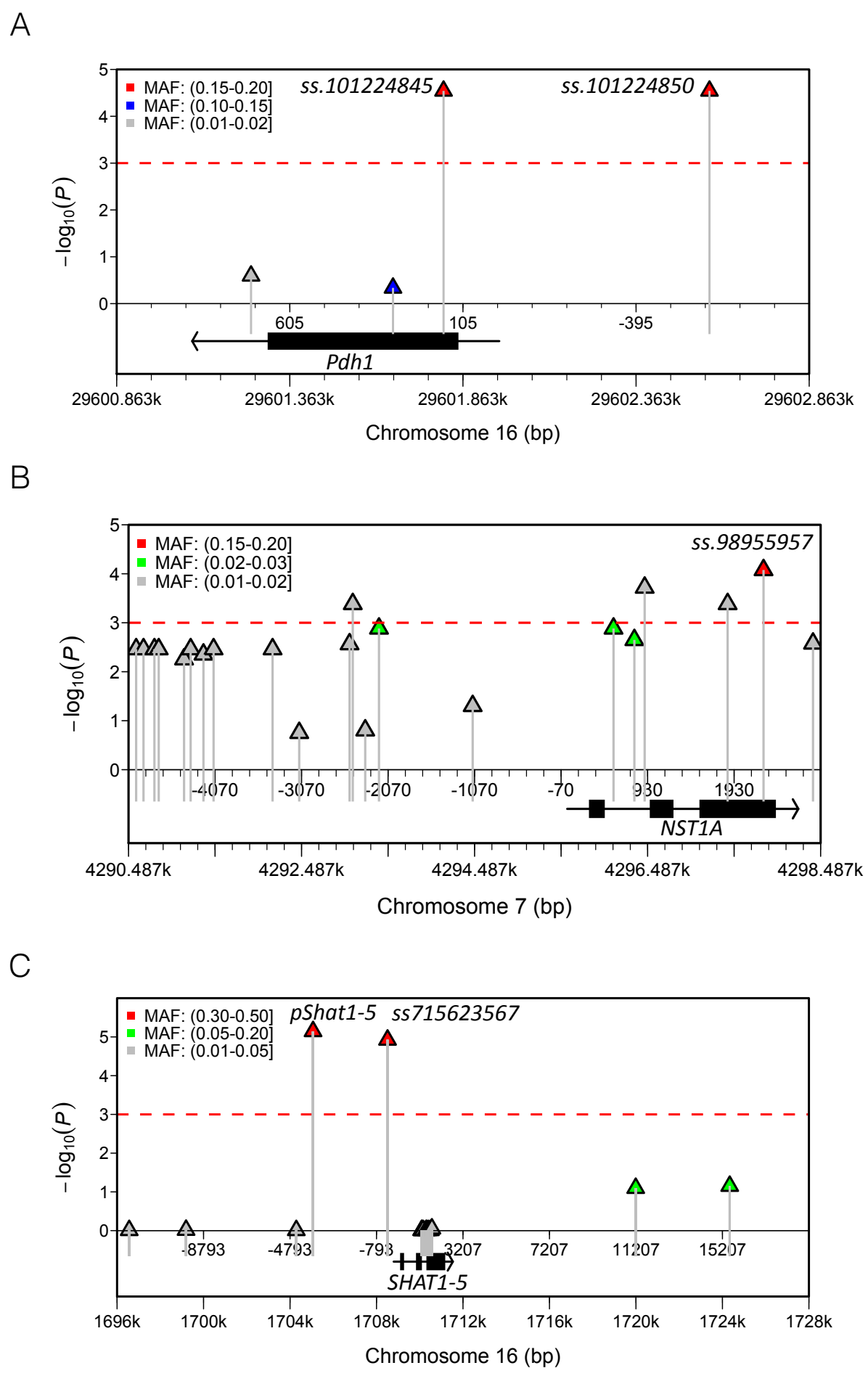

**Figure 7. Candidate gene association analysis for *ss715624201*, *ss715598106* and *ss715623567* associated with pod dehiscence in soybean.** (A), (B) and (C) Regional Manhattan plots of association analysis for *Pdh1*, *NST1A* and *SHAT1-5*, respectively. The x-axis indicates the physical position on the chromosome. Both positions referring to the start of the chromosome and to the transcription start-site are shown. Each triangle represents one SNP. The red dashed lines indicate the significance threshold of $P = 10^{-3}$. The black boxes represent the exons of the gene. The names of the lead SNPs are also given, and their details can be found at Phytozome (https://phytozome.jgi.doe.gov/pz/portal.html#!info?alias=Org_Gmax). MAF = minor allele frequency.

## Phylogenetic analysis and homologs of shattering genes in peas

The phylogenetic analysis of the top hits from the BLAST search for *Pdh1* and *NST1A* against the non-redundant protein sequences of the entire GenBank at NCBI (https://blast.ncbi.nlm.nih.gov/Blast.cgi) showed that homologs of the shattering genes are widely existed in pea family (Fabaceae) species (Fig. 8). The soybean shattering genes are well grouped with legume crops such as pigeonpea (*Cajanus cajan*), kidney bean (*Phaseolus vulgaris*), mung bean (*Vigna radiate var. radiata*), and adzuki bean (*V. angularis*). Additionally, an *NST1* ortholog and a *Pdh1* paralogue were discovered in oilseed rape (*Brassica napus*) and soybean, respectively.

## DISCUSSION

The present study identified multi-level of resistance to pod dehiscence in soybean that was driven by humidity, and was determined by the interactions among QTL harboring *Pdh1*, *NST1A* and *SHAT1-5*, especially between *Pdh1* and *NST1A* loci. Similar phenomenon has been observed in rice. During the rice domestication, *SH4* and *qSH1* fulfilled key roles in the loss of seed shattering. The K79N mutation in SH4, which determines shattering resistance, was prevalent in cultivated rice subspecies *japonica* and *indica* (Lin *et al.* 2007), whereas non-shattering *qSH1* was only found in temperate *japonica* but not tropical *japonica* (subgroup of *japonica*) or *indica* (Konishi *et al.* 2006). Although the mechanism of nonshattering is different between soybean and rice (Dong *et al.* 2014), the underlying driving force of the distribution of the nonshattering alleles might be similar, because the temperate *japonica* mainly grow in cooler zones of the subtropics, temperate zones or in high latitude regions including NEC, Koreas and Japan with lower humidity comparing to the tropical regions (Sweeney *et al.* 2007).

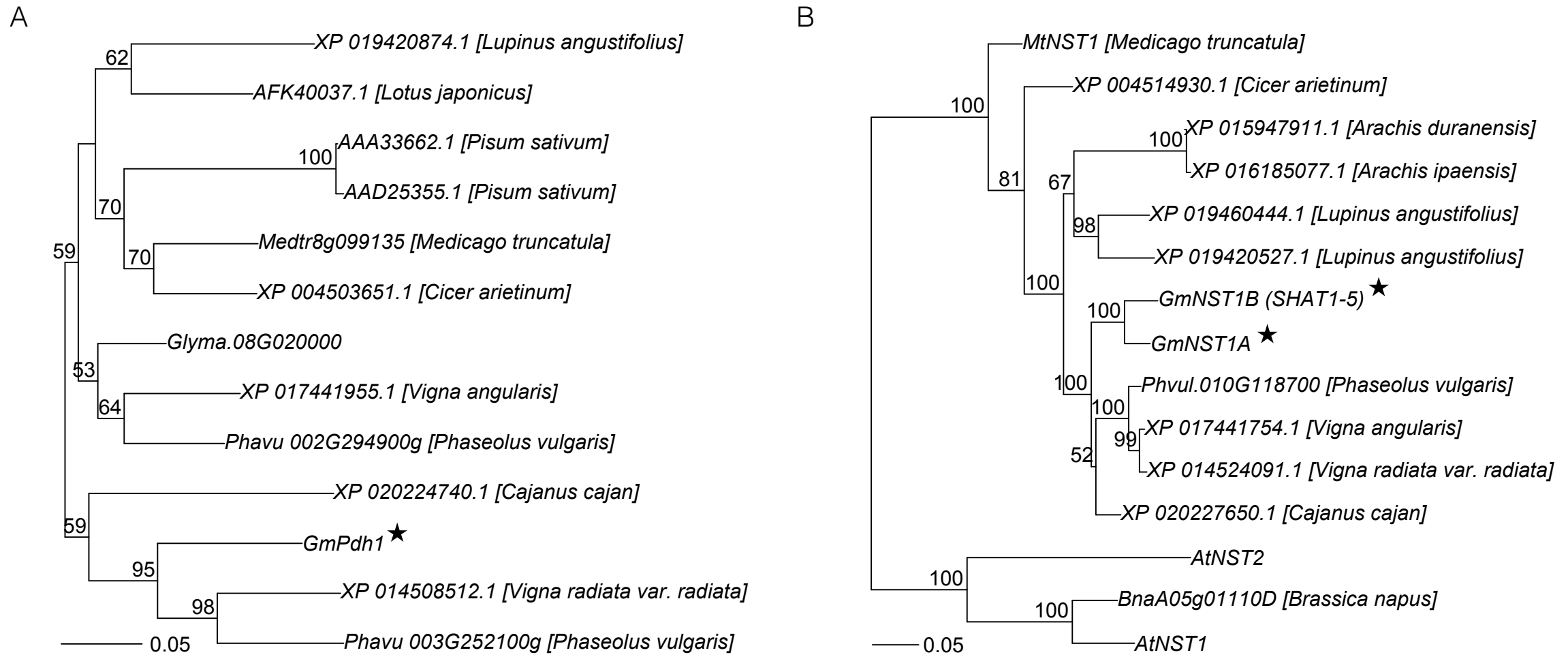

**Figure 8. Neighbor-joining phylogenetic trees of *Pdh1*, *NST1A* and *SHAT1-5* homologs in legume crops**. (A) Phylogenetic tree of *Pdh1* orthologs. (B) Phylogenetic tree of *NST1A* and *SHAT1-5* orthologs. The homologs of *Arabidopsis* and oilseed rape (*Brassica napus*) are also involved. Bootstrap (with 1000 replicates) values with support > 50% are shown. The genes with star are the shattering genes identified in soybean.

Our study provides novel insights into the origin of cultivated soybean from a view of domestication condition. It is generally accepted that soybean originated in China (Fukuda 1933, Hymowitz Newell 1981). However, the center of origin of cultivated soybean in China is controversial. To date, there are four major hypotheses of the origin center: NEC, HHH valleys, SC, and Multiple origin centers (reviewed by (Zhao Gai 2004). The present study suggested that the domestication of soybean in NEC requires, at least, indehiscence at both *Pdh1* and *NST1A*. However, the local wild progenitors had no indehiscent *Pdh1*. This falsifies the hypothesis of NEC origin center. In contrast, our results indicated that HHH valleys was one of the domestication centers, and the cultivated soybean in NEC was disseminated from HHH valleys where indehiscent *Pdh1* originated. Recent studies on genetic diversity among domesticated and wild soybeans also supported the origin center of HHH valleys (Li *et al.* 2008, Li *et al.* 2010, Han *et al.* 2016, Wang *et al.* 2016). Additionally, our study showed that distinct from northern regions, the cultivated soybean in SC requires less or no demand for indehiscent *Pdh1* lending further evidence that SC might be another origin center of the cultivated soybeans (Gai *et al.* 1999, Guo *et al.* 2010).

This study also shed light on the expansion of soybean from China to other regions in Asian. Soybean was first disseminated to peninsular Korea from NEC, and was then introduced into Japan by the sixth century through two paths: i) from NEC through Korea; ii) directly from SC (Singh Hymowitz 1999, Zhao Gai 2004). However, the contribution of each path is unclear. Our results showed that the allele constitution of *Pdh1* among the cultivated soybean in both Korea Peninsula and Japan are substantially different from that in NEC, but similar to that in SC and the coast region in HHH valleys. These observations imply that the cultivated soybean in Korea and Japan are the primarily from SC and/or HHH valleys; and the NEC accounts for very limit contribution, although it geographically connects or is proximal to the above regions. However, this result does not necessary exclude the possibility of the development of cultivated soybean from the local wild progenitors.

The wild soybeans investigated in this study were mainly from Japan, peninsular Korea and NC, but not SC (Fig. 4A). Considering the relative high humidity in SC, the natural selection pressure against the indehiscent *Pdh1* in wild soybean at SC could be even larger than that in HHH valleys. This prompts our speculation that the major-effect indehiscent *Pdh1* among SC soybean landraces was more likely from the expansion of soybeans in HHH valleys rather than originated from the local wild progenitors. However, the natural selection pressure against indehiscent *NST1A* might be very limited in SC. Japan has humidity condition similar to SC, and the indehiscent *NST1A* was predominant among the local wild soybeans (Fig. 6A and Fig. 4B). Therefore, the prevalent indehiscent *NST1A* in SC landraces could be domesticated from the local wild progenitors. A survey of a representative collection of the SC wild progenitors will help uncover the allele constitution of the two pod dehiscent related loci.

Legumes such as soybean and common beans (*Phaseolus vulgarix*) are crops of economic and societal importance, and are the major source of protein, oil and essential nutrients (Schmutz *et al.* 2010, Schmutz *et al.* 2014). A recent study identified that the domain and the function of the *NST1s* were highly conversed between soybean and the model species *Arabidopsis* (Dong *et al.* 2013, Dong *et al.* 2014). Our results showed that a group of legume *NST1A* and *Pdh1* homologs are closely related to each other, indicating that legume crops might also undergone parallel domestication of pod dehiscence as identified recently in cereal species (Lin *et al.* 2012), and the soybean homologs could be used to fasten the domestication of wild legume species that have potential to be economic crops. Current information and studies on legume species are very limited. Further genetic study and genome sequencing of legume crops will help unravel above inference.

The mutation at the promoter region of *SHAT1-5* (designated as *pShat1-5* in the present study) was characterized responsible for pod indehiscence in a recent study (Dong *et al.* 2014). The authors also suggested that *SHAT1-5* was under higher selection intensity than *NST1A* during soybean domestication by comparing the 3’ flanking sequence of *SHAT1*-5 and *NST1A* between 13 wild progenitors and 23 cultivated soybeans. The present study investigated the genetic diversity at related loci using the entire wild and Asian landrace accessions in the USDA soybeans germplasm collection, and illustrated that *SHAT1-5* locus was subjected to higher natural selection pressure than *NST1A* locus. However, the *pShat1-5* was not under extensive selection for pod indehiscence during soybean domestication. In contrast, our results suggested that *NST1A* locus played a critical role in strengthening pod indehiscence in cultivated soybean. The discrepancy between of the present and previous reports (Dong *et al.* 2014) might be attributed to the distinct soybean materials and sample size involved in the two studies.

Previous studies documented two distinct mechanisms of resistance to pod shattering in soybean: i) loss-of-function of *Phd1* decreases the coiling of drying pod walls (Funatsuki *et al.* 2014), and ii) up-regulating the expression of *SHAT1*-5 thickens the FCC secondary walls and strengthen the binding between tow pod walls (Dong *et al.* 2014). However, unlike the paralogue *SHAT1-5*, the present study identified a premature stop codon in *NST1A* associated with pod indehiscence, which is similar to *Pdh1*. We noted that the premature stop codon of *Pdh1* leading to malfunction is near the N terminal of the protein (Fig. 7B), while that of *NST1A* resulting in 47 of 446 amino acids missing is close to the C terminal and the conserved NAC domain at the N terminal remains intact (Fig. 7A) (Dong *et al.* 2013). Interestingly, premature stop in *FSQ6*, a NAC-domain containing transcript factor, has been reported to be responsible for gain-of-function phenotype in *Arabidopsis* (Li *et al.* 2011). Therefore, unlike the truncated *Pdh1* that is a loss-of-function mutation, the truncated *NST1A* might be a gain-of-function mutation, suggesting a third mechanism of pod indehiscence in soybean. Further functional validation of the causal genetic variant of *NST1A* is necessary to uncover the underlying mechanism.

In conclusion, this study demonstrated the importance of the epistatic interaction among three pod dehiscent loci, especially *NST1A* and *Pdh1*, to determine pod-dehiscence during soybean domestication and revealed relative humidity as the driving force in shaping the geographic distribution of indehiscent alleles. It enhanced our understanding about pod dehiscence in soybean by emphasizing the adaptation of genetic control to specific climate conditions. Our results also suggest that the HHH valleys, but not NEC, is the center or at least one of the centers of origin of cultivated soybean. The genetic variants and the correlation between humidity and pod dehiscence identified in this study are valuable for development of resilient soybean cultivars in coping with climate change and quick domestication of the wild legume species.

## ACKNOWLEDGEMENTS

Research funding support from Monsanto Chair in Soybean Breeding and R F Baker Center for Plant Breeding at Iowa State University is sincerely appreciated. This project was also supported by USDA-CRIS IOW04314 project (Crop Genetic Improvement and Adaptation Using Gene Discovery, Phenotypic Prediction, and Systems Engineering). We thank Dr. R V Chowda Reddy for wet lab experiments on candidate gene association analysis and Ms. J. Hicks for proof reading the manuscript.

## AUTHOR CONTRIBUTION

JZ and AKS conceptualized and designed the experiments and research; JZ performed the database searches and statistical analysis; JZ and AKS wrote the manuscript.

**Supporting information:**

Title: Genetic control and geo-climate adaption of pod dehiscence provide novel insights into the soybean domestication and expansion

**Authors**: Jiaoping Zhang* and Asheesh K. Singh*

**Address**: Department of Agronomy, Iowa State University, Ames, IA 50011, USA

* **Correspondence**: singhak@iastate.edu, jiaoping@iastate.edu

**Running title:** Genetics and climate-adaptation of soybean shattering

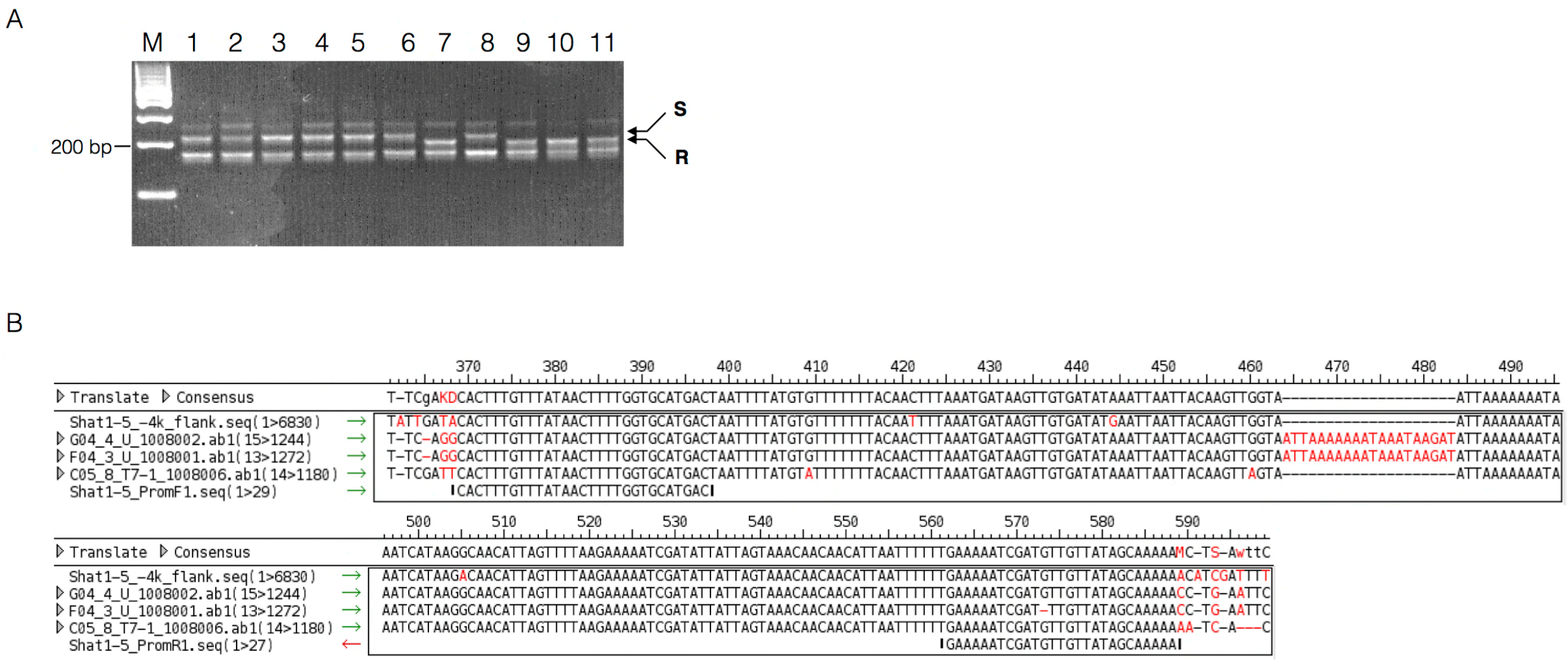

Figure S1. Genotyping and sequencing validation of the known causal polymorphism at the *pShat1-5* region. (A) Shown are the polymorphisms of PCR products of *pShat1-5* region among 11 soybean germplasm accessions. (B) Alignment of the PCR products sequencing results of three accessions and primers against the reference sequence of the soybean cultivar William 82 at the *pShat1-5* region.

Table S1. Primers for the known casual pod dehiscence mutation at the promoter of *SHAT1-5* (*pShat1-5*).

| Primer | 5’ -> 3’ | PCR conditions |
|---|---|---|
| pShat1-5 Forward | CACTTTGTTTATAACTTTTGGTGCATGAC | 94°C 2min; then 35 cycles of 94°C for 20 sec, 58°C for 20 sec, 72°C for 40 sec; and finally 5 min at 72°C. |
| pShat1-5 Reverse | TTTTTGCTATAACAACATCGATTTTTC | |

Table S2 Epistasis test among ss715624201, ss715598106 and ss715623567

| | Df | Sum Sq | Mean Sq | F value | Pr(>F) |
|---|---|---|---|---|---|
| ss715624201 | 1 | 161.2 | 161.17 | 156.7 | < 2e-16 *** |
| ss715598106 | 1 | 15.0 | 14.99 | 14.6 | 0.0001*** |
| ss715623567 | 1 | 5.4 | 5.44 | 5.3 | 0.0217* |
| ss715624201:ss715598106 | 1 | 19.8 | 19.83 | 19.3 | 1.29e-05 *** |
| ss715624201:ss715623567 | 1 | 4.6 | 4.57 | 4.4 | 0.0354* |
| ss715598106:ss715623567 | 1 | 5.2 | 5.18 | 5.0 | 0.0251* |
| ss715624201:ss715598106: ss715623567 | 1 | 0.3 | 0.29 | 0.3 | 0.5973 |
| Residuals | 770 | 792.1 | 1.03 | | |

Significance level: ***, P < 0.001; **, P < 0.01; *, P < 0.05

Table S3 Alleles of the three loci associated with pod dehiscence among 17 ancestors that contributed 86% parentage of North America soybean cultivars.

| Ancestors | ss715624201 | ss715598106 | ss715623567 |
|---|---|---|---|
| FC 033243 | T | G | C |
| PI 548488 | T | G | C |
| PI 548298 | T | T | C |
| PI 548318 | C | G | C |
| PI 548391 | T | G | C |
| PI 548311 | T | G | C |
| PI 548406 | T | G | C |
| PI 548485 | T | G | T |
| PI 548603 | T | G | C |
| PI 548456 | C | G | C |
| PI 548657 | T | G | C |
| PI 548382 | C | G | C |
| PI 548348 | T | T | C |
| PI 548362 | T | G | C |
| PI 548379 | T | G | T |
| PI 548477 | T | G | C |
| PI 548445 | C | G | C |
| Dehiscent allele% | 24% | 12% | 88% |

Table S4 Information of the six of 758 *Glycine soja* accessions carrying the indehiscent allele '*T*' at *ss715624201.*

| Accession | *ss715624201* | Maturity Group | Origin (China) |
|---|---|---|---|
| PI 464938 | *T* | V | Jiangsu |
| PI 468397A* | *T* | IV | Shanxi |
| PI 468397B* | *T* | IV | Shanxi |
| PI 483071A | *T* | IV | Shandong |
| PI 483462B | *T* | IV | Beijing |
| PI 549048 | *T* | III | Beijing |

*Completely identical at the known SNPs of the SoySNP50K data set between the two accessions.